\documentclass[11pt]{article}
\pdfoutput=1

\usepackage{epsfig}
\usepackage{graphicx}
\usepackage{epstopdf}

\usepackage{amsfonts}

\usepackage{amsmath}                                                    
\usepackage{amsthm}                                                     
\usepackage{amssymb}
\usepackage{multirow}
\numberwithin{equation}{section} 
\newcommand{\newc}{\newcommand}
\newc{\be}{\begin{equation}}
\newc{\ee}{\end{equation}}
\newc{\bg}{\begin{gathered}}
\newc{\eg}{\end{gathered}}
\newc{\tref}[1]{Table \ref{#1}}
\newc{\eref}[1]{Equation \eqref{#1}}
\newc{\su}[1]{$SU(#1)$}
\newc{\bm}[1]{\mathbf{#1}}
\newc{\fref}[1]{Figure \ref{#1}}
\newtheorem{theorem}{Theorem}

\newcommand{\eq}[1]{\begin{equation}#1\end{equation}}
\newcommand{\al}[1]{\begin{align}#1\end{align}}

\begin{document}
\begin{titlepage}

\vspace*{0.7cm}

\begin{center}
{
\bf\LARGE
MSSM from F-theory SU(5) with Klein Monodromy }
\\[12mm]
Miguel~Crispim~Rom\~ao,$^{\star}$
\footnote{E-mail: \texttt{m.crispim-romao@soton.ac.uk}}
Athanasios~Karozas$^{\dagger}$
\footnote{E-mail: \texttt{akarozas@cc.uoi.gr}},
Stephen~F.~King$^{\star}$
\footnote{E-mail: \texttt{king@soton.ac.uk}},
George~K.~Leontaris$^{\dagger}$
\footnote{E-mail: \texttt{leonta@uoi.gr}},
Andrew~K.~Meadowcroft$^{\star}$
\footnote{E-mail: \texttt{a.meadowcroft@soton.ac.uk}}
\\[-2mm]

\end{center}
\vspace*{0.50cm}
\centerline{$^{\star}$ \it
Physics and Astronomy, University of Southampton,}
\centerline{\it
SO17 1BJ Southampton, United Kingdom }
\vspace*{0.2cm}
\centerline{$^{\dagger}$ \it
Physics Department, Theory Division, Ioannina University,}
\centerline{\it
GR-45110 Ioannina, Greece}
\vspace*{1.20cm}

\begin{abstract}
\noindent
We revisit a class of $SU(5)$ SUSY GUT models which 
arise in the context of the spectral cover with 
Klein Group monodromy  $V_4=Z_2\times Z_2$.
We show that $Z_2$ matter parities
can be realised via new geometric symmetries respected
by the spectral cover. We discuss a particular example of this kind,
where the low energy effective theory below the GUT scale is just the MSSM with no exotics and 
standard matter parity, extended by the seesaw mechanism with two right-handed neutrinos.
  \end{abstract}

  \end{titlepage}

\thispagestyle{empty}
\vfill
\newpage


\section{Introduction }

Over the last decades  string  theory GUTs  have aroused considerable  interest.
Recent progress has been focused in  F-theory~\cite{Vafa:1996xn,Morrison:1996pp} 
 effective  models~\cite{Beasley:2008dc}-\cite{Blumenhagen:2009yv} which incorporate  
 several constraints  attributed to the topological   properties of the compactified space. 
Indeed, in this context  the  gauge symmetries are associated to the singularities of the 
elliptically fibred compactification manifold.  As such, GUT  symmetries are obtained 
as a subgroup of $E_8$ and the matter content emerges from the decomposition of
the $E_8$-adjoint representation~({for reviews see~\cite{Heckman:2010bq}}).

As is well known, GUT symmetries, have   several interesting features
such as the unification of gauge couplings and the accommodation of fermions
in simple representations.  Yet, they fail to explain the fermion mass  
hierarchy
and more generally, to impose sufficient constraints on the  
superpotential terms. Hence, depending
on the specific model, several rare processes -including proton decay-
are not adequately suppressed. We may infer that, a realistic description of
the observed low energy physics world, requires the existence of
additional symmetry structure of the effective model, beyond the simple GUT
group.

 Experimental observations on limits regarding  exotic processes
(such as baryon and lepton number  as well as  flavour violating cases)
and in particular neutrino physics seem to be nicely explained when
the Standard Model or certain GUTs are extended to include abelian
and discrete symmetries. On purely phenomenological grounds,  $U(1)$  
as well as non-abelian discrete symmetries such as $A_n,S_n, SLP_2(n)$
and so on,  have already been successfully implemented.
However, in this context there is no principle to single out the  
family symmetry group
from the enormous number of possible finite groups.
Moreover, the choice of the scalar spectrum and the
Higgs vev alignments introduce another source of arbitrariness in the models.

In contrast to the above picture, F-theory constructions offer an  
interesting framework for restricting  both the gauge (GUT and discrete) symmetries 
 as well as the available Higgs sector. In the elliptic fibration we end up 
 with an 8-dimensional theory  with a  gauge group  of ADE type. In this work we will focus in the
simplest unified symmetry which is $SU(5)$ GUT. In the present geometric picture,
the $SU(5)$ GUT  is supported by 7-branes wrapping an appropriate  
(del Pezzo) surface $S$ on the internal manifold, while the  number of chiral 
states  is given in terms of a topological index  formula. Moreover, there is no 
use  of adjoint Higgs representations since the breaking down to the 
Standard Model symmetry can occur by turning on a   
non-trivial $U(1)_Y$ flux along the hypercharge generator~\cite{Beasley:2008kw}.
At the same time this mechanism determines   exactly the Standard  
Model matter content.
Further, if the flux parameters are   judiciously chosen
they may provide a solution to the well known doublet triplet splitting
problem of the Higgs sector. In short, in F-theory one can in  
principle develop
all those necessary  tools to determine the GUT group
and predict the matter content of the effective theory.

In the present  work  we will revisit a class of $SU(5)$ SUSY GUT models  which 
arise in the context of the spectral cover.  The reason is that  the recent  
developments in F-theory provide now a clearer insight and a better perspective 
of these constructions. For example, developments on computations  
of the Yukawa couplings\cite{Heckman:2008qa}-\cite{Carta:2015eoh} have shown that a 
reasonable mass hierarchy and mixing may arise
even if more than one  of the fermion families  reside on the same matter curve.  
This implies that effective models left over with only a few matter curves after certain 
monodromy identifications could be viable and it would  be worth  reconsidering them.
More specifically, we will consider the case of the
Klein Group  monodromy  $V_4=Z_2\times Z_2$~\cite{Heckman:2009mn,Dudas:2010zb,Marsano:2011hv,Antoniadis:2012yk}. 
Interestingly, with this particular spectral cover, there are two main ways to implement
its monodromy action, depending on whether $V_4$ is a transitive or  
non-transitive subgroup of $S_4$. A significant part of the present work
will be devoted to the viability of the corresponding  two kinds of effective models.
Another ingredient related to the predictability of the model, is the implementation 
of R-parity conservation, or equivalently a $Z_2$ Matter Parity, which
can be realised with the introduction of new  geometric symmetries~\cite{Hayashi:2009bt} respected
from the spectral cover. In view of these interesting features, 
we also investigate in more detail the superpotential, computing higher  
non-renormalisable corrections, analysing the D and F-flatness conditions and so on.

 The paper is organised as follows. In section 2  we give a short description
 of the derivation of $SU(5)$ GUT  in the context of F-theory.
In section 3 we  describe the action of monodromies  and their role in model building. 
We further focus on the Klein Group monodromy and the corresponding spectral cover
factorisations  which is our main concern in the present  work. In section 4 
we review a few well known mathematical results and theorems which will be used 
in model building of the subsequent sections. 
In section 5 we discuss effective field theory models with Klein Group monodromy
and implement the idea of matter parity of geometric origin. Section 6 deals 
with the particle spectrum, the Yukawa sector and other properties and predictions
of the effective standard model obtained from the above analysis. Finally
we present our conclusions in section 7.

\section{The origin of SU(5) in F-theory}

In this section we explain the  main setup of  these class of models.
Focusing in the  case under consideration, i.e.  the GUT  $SU(5)$,
the effective four dimensional model can be reached from the maximal $E_8$ 
gauge symmetry  through the  decomposition
\[ E_8 \supset SU(5)_{GUT}\times SU(5)_{\perp} \]
In the elliptic fibration,  we know that an ${ SU(5)}$ singularity is described by the
Tate  equation
\be 
y^2=x^3+b_0 z^5+ b_2 xz^3+b_3yz^2+b_4 x^2z+b_5xy
\ee 
 where the homologies of the coefficients in the above equation  are given by:
 \begin{align*}
 [b_k]&=\eta-kc_1\\
 \eta&=6c_1-t
 \end{align*} 
where $c_1$ and $t$ are the Chern classes of the Tangent and Normal  
 bundles respectively.

The first $SU(5)$ is defining the GUT group of the effective theory,
the second  $SU(5)_{\perp}$ incorporates additional symmetries  
of the effective theory while it can be described in the context of the spectral cover.
Indeed, in this picture, one can depict the non-abelian Higgs bundle
in terms of the adjoint scalar  field configuration~\cite{Donagi:2009ra}
and work with the Higgs eigenvalues and eigenvectors.
For $SU(n)$ these emerge as roots of a characteristic polynomial
of $n$-th degree. Thus the $SU(5)$ spectral surface $C_5$ is represented
  by the  fifth order polynomial
   \be
   { C_5}=b_0s^5+b_1s^4+b_2s^3+b_3s^2+b_4s+b_5
   \,=\,b_0 
   \prod^{5}_{i=1}(s-t_i)\label{sc10}
   \ee
Since the roots are associated to the $SU(5)$ Cartan subalgebra
their sum is zero, $\sum_it_i=0$,  thus we  have put $b_1=0$.

The $5+\bar 5$ and $10+\overline{10} $  representations are found at certain ehnancements
of the $SU(5)$ singularity. In particular, for this purpose  
 the relevant quantities are~\cite{Donagi:2009ra}
 \begin{eqnarray} 
{\cal P}_{10} &=& b_5=\prod_i t_i\\
 {\cal P}_5&=& b_3^2b_4-b_2b_3b_5+b_0b_5^2\propto \prod_{i\neq j} (t_i+t_j)
 \end{eqnarray}
At the ${\cal P}_{10}=0$ locus the enhanced singularity is $SO(10)$  and 
 the intersection defines the matter curve accommodating the $10$'s.
 Fiveplets are found at a matter curve defined at an  $SU(6)$ enhancement 
 associated to the locus ${\cal P}_{5}=0$.

 In practive, we are interested in phenomenologically 
viable cases where the spectral cover splits in several pieces.
Consider for example the  splitting 
expressed through the  breaking chain
\[ E_8 \to SU(5)\times SU(5) \to SU(5) \times U(1)^4 \]
where we assumed breaking of $SU(5)_{\perp}$ along the Cartan, $\sum_it_i=0$.
The presence of four $U(1)$'s in the effective theory leaves no room for a viable
 superpotential, since many of the required terms, including the top Yukawa coupling, 
 are not allowed. Nevertheless, monodromies imply various kinds of symmetries among the
roots $t_i$ of the spectral cover polynomial which can be used to relax
these tight constraints. The particular relations among these roots
depend on the details of the compactification
and the geometrical properties of the internal manifold.
All possible ways fall into some Galois group which in the case of  
$SU(5)_{\perp}$
is a subgroup of the corresponding Weyl group,  i.e., the group $S_5$
of all possible permutations of the five Cartan weights $t_i$.
It is obvious that there are several  options  and each of them leads to models with 
completely different properties and predictions of the
effective field theory. 
 Before starting our investigations on
the effective models derived in the context of the aforementioned 
monodromy,  we will analyse these issues in
the next section.

\newpage
\section{The Importance of Monodromy}

  For the $SU(5)_{GUT}$ model, we have seen that  any possible remnant  
symmetries
  (embedable in the $E_8$ singularity)  must be contained in $SU(5)_{\perp}$.
We have already explained that in the spectral cover approach
  we  quotient the theory by the action of a finite group~\cite{Heckman:2009mn}
which is expected to descend from a geometrical symmetry of the  
compactification.
Starting form an  $C_5$ spectral cover, the local field theory
  is determined by the $SU(5)$ GUT group and  the Cartan subalgebra of  
  $SU(5)_{\perp}$
modulo the Weyl group $W({\tiny SU(5)_{\perp}})$. This is the group  
$S_5$, the permutation symmetry
of five elements which in the present case correspond to the Cartan   
weights $t_{1,\dots 5}$.

 Depending on the geometry of the manifold, $C_5$ may slit to several factors
\[  C_5 =\prod_j C_j\]
For the present work, we will assume two cases where the  
compactification geometry
implies the splitting of the spectral cover to  $C_5\to C_4\times C_1$  
and $C_5\to C_2\times C_2'\times C_1$.
Assuming the splitting  $C_5\to C_4\times C_1$, the permutation takes  
place between the four roots, say $t_{1,2,3,4}$,
and the corresponding Weyl group is $S_4$. Notwithstanding, under  
specific geometries to be discussed in the subsequent sections,
  the monodromy may be described by the Klein group $V_4\in S_4$. The  
latter might be either transitive or non transitive.
  This second case implies the spectral cover factorisation $C_4\to  
C_2\times C_2'$.
As a result, there are two non-trivial identifications   acting on the  
pairs $(t_1,t_2)$ and $(t_3,t_4)$ respectively while
both are described by the Weyl group $W(SU(2)_{\perp})\sim S_2 $.    
Since $S_2\sim Z_2$, we conclude that
in this case the monodromy action is the non-transitive Klein group  
$Z_2\times Z_2$.
Next, we will analyse the basic features of these two spectral cover  
factorisations.

\subsection{$S_4$ Subgroups and Monodromy Actions}

\begin{figure}[t!]\centering
\includegraphics[scale=0.8]{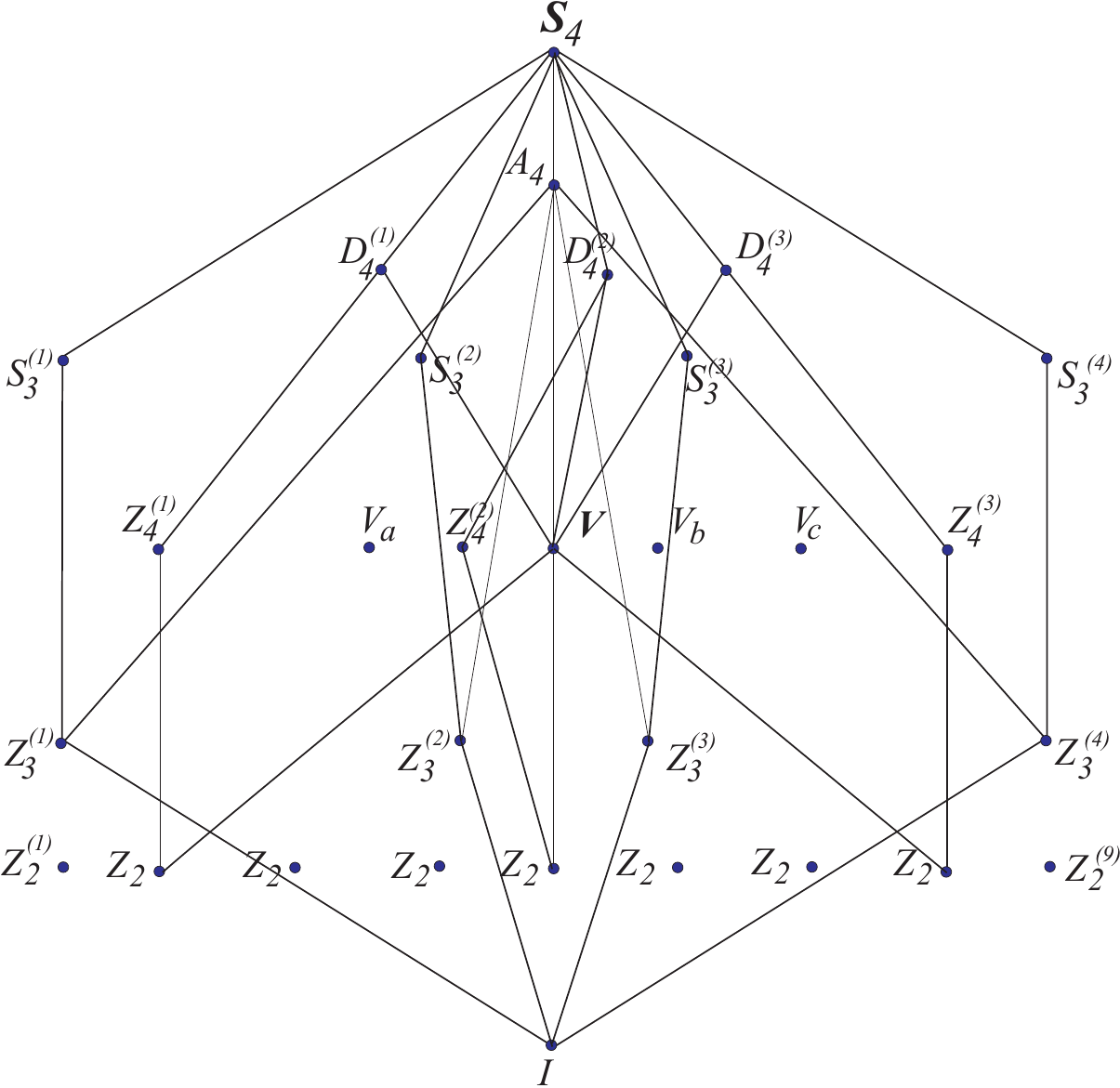}
\caption{Pictorial summary of the subgroups of $S_4$, the group of all
 permutations of four elements - representative of the symmetries of a cube.\label{spider}}
\end{figure}

The group of all permutations of four elements, $S_4$,  has a total of  
$24$ elements.\footnote{The order of an $S_N$ group is given by $N!$}  
These include 2,3,4 and 2+2-cycles, all of which are listed in  
\tref{cycles}. These cycles form a total of 30 subgroups of $S_4$,  
shown in \fref{spider}. Of these there are those subgroups that are  
transitive subgroups of $S_4$: the whole group, $A_4$, $D_4$, $Z_4$  
and the Klein group.

We focus now  in compactification geometries consistent  with  the  
Klein group monodromy  $V_4=Z_2\times Z_2$.
We observe that there are three non-transitive $V_4$ subgroups within  
$S_4$ and only one transitive  subgroup. This transitive Klein group  
is the subgroup of the $A_4$ subgroup. Considering \tref{cycles}, one  
can see that $A_4$ is the group of all even permutations of four  
elements and the transitive $V_4$ is that group excluding 3-cycles.  
The significance of this is that in the case of Galois theory, to be  
discussed in Section \ref{galois theory}, the transitive subgroups  
$A_4$ and $V_4$ are necessarily irreducible quartic polynomials, while  
the non-transitive $V_4$ subgroups of $S_4$ should be reducible.

\begin{table}[t!]\centering
\begin{tabular}{|c|c|c|c|}\hline
&$S_4$ cycles & Transitive $A_4$ & Transitive $V_4$\\\hline
4-cycles& $(1234)$, $(1243)$, $(1324)$, $(1342)$, $(1423)$, $(1432)$ &  
No & No\\
3-cycles& $(123)$, $(124)$, $(132)$, $(134)$, $(142)$, $(143)$,  
$(234)$, $(243)$ & Yes& No\\
2+2-cycles& $(12)(34)$, $(13)(24)$, $(14)(23)$ &Yes & Yes \\
2-cycles& $(12)$, $(13)$, $(14)$, $(23)$, $(24)$, $(34)$ & No & No\\
1-cycles& $e$ &Yes&Yes\\\hline
\end{tabular}
\caption{A summary of the permutation cycles of $S_4$, categorised by cycle size and whether or not those cycles are contained within the transitive subgroups $A_4$ and $V_4$. This also shows that $V_4$ is necessarily a transitive subgroup of $A_4$, since it contains all the $2+2$-cycles of $A_4$ and the identity only. \label{cycles}}
\end{table}

In terms of group elements, the Klein group that is transitive in  
$S_4$ has the elements:
\be\{(1), (12)(34), (13)(24), (14)(23)\} \ee
which are the 2+2-cycles shown in \tref{cycles} along with the  
identity. On the other hand, the non-transitive Klein groups within  
$S_4$ are isomorphic to the subgroup containing the elements:
\be V_4=\{(1),(12),(34),(12)(34)\}\ee
The distinction here is that the group elements are not all within one  
cycle, since we have two 2-cycles and one 2+2-cycle. These types of  
subgroup must lead to a factorisation of the quartic polynomial, as we  
shall discuss in Section \ref{galois theory}. Referring to  
\fref{spider}, these Klein groups are the nodes disconnected from the  
web, while the central $V_4$ is the transitive group.

\subsection{  Spectral cover factorisation}

In this section we will discuss the two possible factorisations 
of the spectral surface compatible with a Klein Group monodromy,
in accordance with the previous analysis.
In particular, we shall be examining the implications of a monodromy  
action  that is a subgroup of $S_4$ - the most general monodromy  
action relating four weights. In particular we shall be interested in  
the chain of subgroups $S_4\rightarrow A_4 \rightarrow V_4$, which we  
shall treat as a problem in Galois theory. \\

\subsubsection{${\mathcal C_4}$ spectral cover}

This set of monodromy actions require the spectral cover of  
\eref{sc10} to split into a linear part and a quartic part:
\begin{align}{\mathcal C_5}&\rightarrow {\mathcal C_4\times C_1}\\
{\mathcal C_5}& \rightarrow  
(a_5s^4+a_4s^3+a_3s^2+a_2s+a_1)(a_6+a_7s)\label{sc10f}
\end{align}
The $b_1=0$ condition must be enforced for \su{5} tracelessness. This  
can be solved by consistency in \eref{sc10f},
\begin{equation}
b_1=a_5a_6+a_4a_7=0\,.
\end{equation}
Let us introduce a new section $a_0$, enabling one to write a general  
solution of the form:
\begin{align*}\label{trace}
a_4&=\pm a_0a_6\\
a_5&=\mp a_0a_7
\end{align*}
Upon making this substitution, the defining equations for the matter  
curves are:
\begin{align}
{\mathcal C_5\,:}&=a_1a_6\\
{\mathcal C_{10}\,:}&=(a_2^2a_7+a_2a_3a_6\mp  
a_0a_1a_6^2)(a_3a_6^2+(a_2a_6+a_1a_7)a_7) \label{s4fives}
\end{align}
which is the most general, pertaining to an $S_4$ monodromy action on  
the roots. By consistency between \eref{sc10f} and \eref{sc10}, we can  
calculate that the homologies of the coefficients are:
\begin{align*}
[a_i]&=\eta-(i-6)c_1 -\chi\\
i&=1,2,3,4,5\\
[a_6]&=\chi\\
[a_7]&=c_1+\chi\\
[a_0]&=\eta-2(c_1+\chi)
\end{align*}

\subsubsection{The $C_2\times C_2'\times C_1$ case}

If the $V_4$ actions  are not derived as transitive subgroups of  
$S_4$, then the Klein group is isomorphic to:
\be
A_4 \not\supset V_4:\,\,\{(1),(12),(12)(34),(34)\}\label{V4NT}
\ee
   This is not contained in $A_4$, but is admissible from the spectral  
cover in the form of a monodromy
   $C_5\rightarrow C_2\times C_2'\times C_1$.

Then,  the $ {\bf 10}\in SU(5)$ GUT (${\bf }\in SU(5)_{\perp}$)  
spectral cover reads
\eq{\label{eq:C5}
  C_5 : (a_1  + a_2 s + a_3 s^2)(a_4  + a_5 s +  a_6 s^2)(a_7 +a_8 s)
}
We may now match the coefficients of this polynomial in each order in  
$s$ to the ones of the spectral cover with the $b_k$ coefficients:
\al{
b_0 &= a_{368}\nonumber  \\
b_1 &= a_{367}+a_{358}+a_{268} \nonumber \\
b_2 &= a_{357}+a_{267}+a_{348}+a_{258}+a_{168}  \label{eq:bnamsol}\\
b_3 &= a_{347}+a_{257}+a_{167}+a_{248}+a_{158} \nonumber \\
b_4 &= a_{247}+a_{157}+a_{148} \nonumber \\
b_5 &= a_{147}\nonumber
}
following the notation $a_{ijk}=a_i a_j a_k$ in~\cite{Dudas:2010zb}. In order to find the  
homology classes of the new coefficients $a_i$,  we match the  
coefficients of the above polynomial in each order in $s$ to the ones  
of \eref{sc10} such that we get relations of the form $b_k = b_k(a_i)$.

Comparing to the homologies of the unsplit spectral cover, a  
solution for the above can be found for the homologies of $a_i$.  
Notice, though, that we have 6 well defined homology classes for $b_j$  
with only 8 $a_i$ coefficients, therefore the homologies of $a_i$ are  
defined up to two homology classes:
\al{
[a_{n=1,2,3}]&=\chi_1+(n-3)c_1\nonumber\\
[a_{n=4,5,6}]&=\chi_2+(n-6)c_1\label{eq:anhomol}\\
[a_{n=7,8}]&=\eta+(n-8)c_1-\chi_1-\chi_2\nonumber
}

We have to enforce the $SU(5)$ tracelessness condition, $b_1 = 0 $. An  
Ansatz for the solution was put forward in 
\cite{Dudas:2010zb},
\al{
a_2 &= -c (a_6 a_7 +a_5 a_8) \nonumber  \\
a_3 &= c a_6 a_8 \label{eq:splitsol}
}
which introduces a new section, $c$, whose homology class is  
completely defined by
\eq{
[c]= -\eta + 2 \chi_1
}

With this anstaz for the solution of the splitting of spectral cover,  
$P_{10}$ reads
\eq{
P_{10} = a_1 a_4 a_7
}
while the $P_5$ splits into
\al{
P_5 =& a_5 (a_6 a_7+a_5 a_8) (a_6 a_7^2+a_8 (a_5 a_7+a_4 a_8))(a_1-a_5  
a_7 c) \\
&  (a_1^2-a_1 (a_5 a_7+2 a_4 a_8) c+a_4 (a_6 a_7^2+a_8 (a_5 a_7+a_4  
a_8)) c^2),
}
An extended analysis of this interesting case will be presented in the  
subsequent sections.

\section{A little bit of Galois theory}\label{galois theory}
So far, we have outlined the properties of the most general spectral  
cover with a monodromy action acting on four of the roots of the  
perpendicular \su{5} group. This monodromy action is the Weyl group  
$S_4$, however a subgroup is equally admissible as the action.  
Transitive subgroups are subject to the theorems of Galois theory,  
which will allow us to determine what properties the coefficients of the  
quartic factor of~\eref{sc10f} must have in order to have roots with a  
particular symmetry \cite{Marsano:2009gv}-\cite{Karozas:2015zza}. In this paper we shall focus on the Klein group,  
$V_4\cong Z_2\times Z_2$. As already mentioned, the transitive $V_4$  
subgroup of $S_4$ is contained within the $A_4$ subgroup of $S_4$, and  
so shall share some of the same requirements on the coefficients. \\

While Galois theory is a field with an extensive literature to  
appreciate, in the current work we need only reference a handful of  
key theorems. We shall omit proofs for these theorems as they are  
readily available in the literature and are not required for the  
purpose at hand.
\begin{theorem}\label{thm1}
Let $K$ be a field with characteristic different than $2$, and let  
$f(X)$ be a separable, polynomial in $K(X)$ of degree $n$.
\begin{itemize}
\item
If $f(X)$ is irreducible in $K(X)$ then its Galois group over $K$ has  
order divisible by $n$.
\item
The polynomial $f(X)$ is \emph{irreducible} in $K(X)$ \emph{if and  
only if} its Galois group over $K$ is a transitive subgroup of $S_n$.
\end{itemize}
\end{theorem}
This first theorem offers the key point that any polynomial of degree  
$n$, that has non-degenerate roots, but cannot be factorised into  
polynomials of lower order with coefficients remaining in the same  
field must necessarily have a Galois group relating the roots that is  
$S_n$ or a transitive subgroup thereof.
\begin{theorem}\label{thm2}
Let $K$ be a field with characteristic different than $2$, and  
let $f(X)$ be a separable, polynomial in $K(X)$ of degree $n$. Then  
the Galois group of $f(X)$ over $K$ is a subgroup of $A_n$ \emph{if  
and only if} the discriminant of $f$ is a square in $K$.
\end{theorem}
As already stated, we are interested specifically in transitive $V_4$  
subgroups. Theorem \ref{thm2} gives us the requirement for a Galois  
group that is $A_4$ or its transitive subgroup $V_4$ - both of which  
are transitive in $S_4$. Note that no condition imposed on the  
coefficients of the spectral cover should split the polynomial ($C_4  
\to C_2 \times C_2 $), due to Theorem \ref{thm1}. We also know by  
Theorem \ref{thm2} that both $V_4$ and $A_4$ occur when the  
discriminant of the polynomial is a square, so we necessarily require  
another mechanism to distinguish the two.\\

\subsection{The  Cubic Resolvent}
The so-called Cubic Resolvent, is an expression for a cubic polynomial  
in terms of the roots of the original quartic polynomial we are  
attempting to classify. The roots of the cubic resolvent are defined  
to be,
\be  
x_1=(t_1t_2+t_3t_4),\,x_2=(t_1t_3+t_2t_4),\,x_3=(t_1t_4+t_2t_3) 
\label{crroots}\ee
and one can see that under any permutation of $S_4$ these roots  
transform between one another. However, in the event that the  
polynomial has roots with a Galois group relation that is a subgroup  
of $S_4$, the roots need not all lie within the same orbit. The  
resolvent itself is defined trivially as:
\be (x-(t_1 t_2 + t_3 t_4))(x-(t_1 t_3 + t_1 t_4))(x-(t_1 t_4 + t_3  
t_2)) = g_3 x^3 + g_2 x^2 +g_1 x +g_0 \ee
The coefficients of this equation can be determined by relating of the  
roots to the original ${\mathcal C_4}$ coefficients. This resulting  
polynomial is:
  \be g(x)= a_5^3 x^3-a_3 a_5^2 x^2+\left(a_2 a_4-4 a_1 a_5\right) a_5  
x-a_2^2 a_5+4 a_1 a_3
    a_5-a_1 a_4^2  \label{cres1}\ee
Note that this may be further simplified by making the identification  
$y=a_5x$.
\be g(y) = y^3-a_3 y^2+\left(a_2 a_4-4 a_1 a_5\right) y-a_2^2 a_5+4 a_1 a_3
    a_5-a_1 a_4^2 \label{cres2}\ee
If the cubic resolvent is factorisable in the field $K$, then the  
Galois group does not contain any three cycles. For example, if the  
Galois group is $V_4$, then the roots will transform only under the  
2+2-cycles:
\be
V_4 \subset A_4 = \{(1),(12)(34),(13)(24),(14)(23)\}\,.
\ee
Each of these actions leaves the first of the roots in~\eref{crroots}  
 invariant, thus implying that the cubic resolvent is reducible  
in this case. If the Galois group were $A_4$, the 3-cycles present in  
the group would interchange all three roots, so the cubic resolvent is  
necessarily irreducible. This leads us to a third theorem, which  
classifies all the Galois groups of an irreducible quartic polynomial (see also Table \ref{gal}).
\begin{theorem}\label{thm3}
The Galois group of a quartic polynomial $f(x)\in K$, can be described  
in terms of whether or not the discriminant of $f$ is a square in $K$  
and whether or not the cubic resolvent of $f$ is reducible in $K$. 
\begin{table}[t!]\centering
\begin{tabular}{|c|c|c|}\hline
Group&Discriminant&Cubic Resolvent\\\hline
$S_4$&$\Delta\neq\delta^2$&Irreducible\\
$A_4$&$\Delta=\delta^2$&Irreducible\\
$D_4$/$Z_4$&$\Delta\neq\delta^2$&Reducible\\
$V_4$&$\Delta=\delta^2$&Reducible\\\hline
\end{tabular}\caption{A summary of the conditions on the partially symmetric 
polynomials of the roots and their corresponding Galois group.}\label{gal}
\end{table}
\end{theorem}

\section{Klein monodromy and the origin of matter parity}

In this section we will analyse a class of four-dimensional effective  models obtained 
under the assumption that  the compactification  geometry induces a $Z_2\times Z_2$ monodromy.  
As we have seen in the previous section, there are two distinct ways to realise
this scenario, which depends on whether the corresponding Klein group is  transitive or non-transitive.
In the present work we will choose to explore the rather promising case 
where the monodromy Klein group is  non-transitive.  In other words, this essentially 
means that the spectral  cover admits a $C_2\times C_2'\times C_1$ factorisation.
The case of a transitive Klein group is more involved and it is not easy 
to obtain a viable effective model, hence we will consider this issue
in a future work.

Hence, turning our attention to the non-transitive case, 
the basic structure of the model obtained in this case  corresponds to  
one of those initially presented  in~\cite{Heckman:2009mn} and  
subsequently  elaborated by other authors~\cite{Dudas:2010zb}-\cite{Antoniadis:2012yk}.
This model possesses several phenomenologically interesting features  
and we consider it is worth elaborating it further.

\subsection{Analysis of the $Z_2\times Z_2$ model}

To set the stage, we first present a short review of the basic  
characteristics of the model following mainly the
notation of~\cite{Dudas:2010zb}.
The $Z_2 \times Z_2$ monodromy case implies a $2+2+1$ splitting  
of the spectral fifth-degree polynomial which  has already been given in~(\ref{eq:C5}).
Under the action (\ref{V4NT}), for each element, either $x_2$ and $x_3$ roots 
defined in~(\ref{crroots}) are exchanged or   the roots are unchanged.

The effective model is characterised by  three distinct $10$ matter curves, and five $5$ matter  
curves. The matter curves, along with their charges under the perpendicular surviving $U(1)$
and their homology classes are presented in table \ref{tab:homologycharges}.\\
\begin{table}[h]
\begin{center}
\begin{tabular}{|c|c|c|c|}\hline
          Curve   & $U(1)$  Charge   &                           
Defining Equation                                          &     
Homology Class                  \\ \hline
          $10_1$  &  $t_{1}$   &                                      
$a_1$                                      &         $ - 2 c_1 +  
\chi_1$            \\
          $10_3$  &  $t_{3}$   &                                      
$a_4$                                      &         $ - 2 c_1 +  
\chi_2$            \\
          $10_5$  &   $t_5$    &                                      
$a_7$                                      &     $ \eta -c_1 -\chi_1  
-\chi_2$     \\
         $5_{1}$  &  $-2t_1$   &                               $a_6  
a_7+a_5 a_8$                                &        $\eta - c_1 -  
\chi_1$           \\
         $5_{13}$ & $-t_1-t_3$ & $ a_1^2-a_1 (a_5 a_7+2 a_4 a_8) c+  
a_4 (a_6 a_7^2+a_8 (a_5 a_7+a_4 a_8)) c^2 $ &        $ - 4 c_1 + 2  
\chi_1 $         \\
         $5_{15}$ & $-t_1-t_5$ &                                 
$a_1-a_5 a_7 c$                                 &          $- 2 c_1 +  
\chi_1$           \\
         $5_{35}$ & $-t_3-t_5$ &                       $a_6 a_7^2+a_8  
(a_5 a_7+a_4 a_8) $                       & $2\eta - 2 c_1 - 2 \chi_1  
- \chi_2 $  \\
         $5_{3}$  &  $-2t_3$   &                                      
$a_5$                                      &             
$-c_1+\chi_2$\\\hline
\end{tabular}
\caption{Matter curves and their charges and homology  
classes}\label{tab:homologycharges}
\end{center}
\end{table}

Knowing the homology classes associated to each curve allows us to  
determine the spectrum of the theory through the units of abelian  
fluxes that pierce the matter curves. Namely, by turning on a flux in  
the $U(1)_X$ directions, we can endow our spectrum with chirality and  
break the perpendicular group. In order to retain an anomaly free  
spectrum we need to allow for
\eq{
\sum M_{5} + \sum M_{10} = 0,
}
where $M_5$ ($M_{10}$) denote $U(1)_X$ flux units piercing a certain  
$5$ ($10$) matter curve.

A non-trivial flux can also be turned on along the Hypercharge. This  
will allow us to split GUT irreps, which will provide a solution for  
the doublet-triplet splitting problem. In order for the Hypercharge to  
remain ubroken, the flux configuration should not allow for a  
Green-Schwarz mass, which is accomplished by
\eq{
F_Y \cdot c_1 = 0 , \ F_Y \cdot \eta = 0 .
}

For the new, unspecified, homology classes, $\chi_1$ and $\chi_2$ we  
let the flux units piercing them to be
\eq{
F_Y \cdot \chi_1 = N_1 , \ F_Y \cdot \chi_2 = N_2,
}
where $N_1$ and $N_2$ are flux units, and are free parameters of the theory.

For a fiveplet, $5$ one can use the above construction as a  
\emph{doublet-triplet splitting solution} as
\al{
n(3,1)_{-1/3}-n(\overline{3},1)_{1/3}=M_5 , \\
n(1,2)_{1/2}-n(1,2)_{-1/2}=M_5 + N ,
}
where the states are presented in the SM basis. For a $10$ we have
\al{
n(3,2)_{1/6}-n(\overline{3},2)_{-1/6}=M_{10} , \\
n(\overline{3},1)_{-2/3}-n(3,1)_{2/3}=M_{10} - N , \\
n(1,1)_{1}-n(1,1)_{-1}=M_{10} + N.
}

In the end, given a value for each $M_5$, $M_{10}$, $N_1$, $N_2$ the  
spectrum of the theory is fully defined as can be seen in  
\tref{tab:homologyfluxesspectrum}

\begin{table}[h]
\begin{center}
\begin{tabular}{|c|c|c|c|c|c|}\hline
          Curve   &  Weight   &               Homology               &  
     $N_Y$     &    $N_X$ &                          Spectrum         
\\ \hline
          $10_1$  &  $t_{1}$   &         $ - 2 c_1 + \chi_1$           
&     $N_1$     &  $M_{10_1}$ &       $M_{10_1}  
Q+(M_{10_1}-N_1)u^c+(M_{10_1}+N_1)e^c    $  \\
          $10_3$  &  $t_{3}$   &         $ - 2 c_1 + \chi_2$           
&     $N_2$     &  $M_{10_3}$ &       $M_{10_3}  
Q+(M_{10_3}-N_2)u^c+(M_{10_3}+N_2)e^c    $ \\
          $10_5$  &   $t_5$    &     $ \eta -c_1 -\chi_1 -\chi_2$      
& $-N_1 - N_2$  &  $M_{10_5}$  &   $M_{10_5}  
Q+(M_{10_5}+N)u^c+(M_{10_5}-N)e^c    $  \\
         $5_{1}$  &  $-2t_1$   &        $\eta - c_1 - \chi_1$          
&    $-N_1$     & $M_{5_{1}}$  &       $M_{5_1} \overline{d^c} +  
(M_{5_1}-N_1) \overline{L}$ \\
         $5_{13}$ & $-t_1-t_3$ &        $ - 4 c_1 + 2 \chi_1 $         
&    $2 N_1$    & $M_{5_{13}}$ &    $M_{5_{13}} \overline{d^c} +  
(M_{5_{13}}+2 N_1) \overline{L}$  \\
         $5_{15}$ & $-t_1-t_5$ &          $- 2 c_1 + \chi_1$           
&     $N_1$     & $M_{5_{15}}$  &    $M_{5_{15}} \overline{d^c} +  
(M_{5_{15}}+N_1) \overline{L}$\\
         $5_{35}$ & $-t_3-t_5$ & $2\eta - 2 c_1 - 2 \chi_1 - \chi_2 $  
& $-2 N_1- N_2$ & $M_{5_{35}}$  & $M_{5_{35}} \overline{d^c} +  
(M_{5_{35}}-2 N_1-N_2) \overline{L}$ \\
         $5_{3}$  &  $-2t_3$   &            $-c_1+\chi_2$              
&     $N_2$     & $M_{5_{3}}$ &     $M_{5_{3}} \overline{d^c} +  
(M_{5_{3}}+N_2) \overline{L}$\\\hline
\end{tabular}
\caption{Matter curve spectrum. Note that $N=N_1+N_2$ has been used as  
short hand.\label{tab:homologyfluxesspectrum}}
\end{center}
\end{table}

\subsection{Matter Parity}

It was first proposed before~\cite{Hayashi:2009bt}, in  
local F Theory constructions there are geometric discrete symmetries  
of the spectral cover that manifest on the final field theory. To see  
this note that the spectral cover equation is invariant, up to a phase,  
under the transformation $\sigma: s \mapsto \sigma(s)$ of the  
fibration coordinates, such that
\al{
s &\to s e^{i\phi} \\
b_k & \to b_k e^{i \chi} e^{i(k-6)\phi}.\label{SPS}
}
As detailed in~\cite{Antoniadis:2012yk} this can be associated to 
a symmetry of the  matter fields residing on the various curves.  
We can use the  equations relating $b_k\propto a_l a_m a_n$, with $l+m+n=17$,  
to find the transformation rules of the $a_k$ such that the  
spectral cover equation respects the symmetry~(\ref{SPS}). 
This implies that the coefficients $a_n$ should transform as
\eq{
a_n \to e^{i \psi_n} e^{i(11/3-n)\phi} a_n .
}

We now note that the above transformations can be achieved by a $Z_N$  
symmetry if $\phi = 3 \frac{2 \pi}{N}$. In that case one can find, by  
looking at the equations (\ref{eq:bnamsol}) for $b_k \propto a_l a_m  
a_n$ that we have
\al{
\psi_1 &= \psi_2 = \psi_3 \\
\psi_4 &= \psi_5 =\psi_6 \\
\psi_7 &= \psi_8
}
meaning that there are three distinct cycles, and
\eq{
\chi = \psi_1 + \psi_4 + \psi_7 .
}

Furthermore, the section $c$ introduced to split the matter conditions  
(\ref{eq:splitsol}) has to transform as
\eq{
c \to e^{i \phi_c} c ,
}
with
\eq{
\phi_c = \psi_3 -\psi_6 -\psi_7 + \left( - \frac{11}{3}+11 \right)  
\phi \quad{,}\quad 
\phi_c =  \psi_2 -\psi_5 -\psi_8 + \left( - \frac{11}{3}+11 \right) \phi
}

We can now deduce what would be the matter parity assignments for  
$Z_2$ with $\phi = 3 (2 \pi/2)$. Let $p(x)$ be the parity of a section  
(or products of sections), $x$. We notice that there are relations between  
the parities of different coefficients, for example one can easily find
\eq{
\frac{p(a_1)}{p(a_2)}  = -1
}
amongst others, which allow us to find that all parity assignments  
depend only on three independent parities
\al{
p(a_1) &= i \\
p(a_4) &= j \\
p(a_7) &= k \\
p(c) & = ijk ,
}
where we notice that $i^2=j^2=k^2=+$.
The parities for each matter curve -- both in form of a function of  
$i,j,k$ and all possible assignments -- can  are presented  
in the table \ref{tab:allpar}.

\begin{table}[h]
\begin{center}
\begin{tabular}{|c|c|c|c|c|c|c|c|c|c|c|}\hline
          Curve   &   Charge  & Parity  & \multicolumn{7}{c}{All  
possible assignments}& \\ \hline
          $10_1$  &  $t_{1}$  &  $i$    & $+$ & $-$ & $+$ & $-$ & $+$  
& $-$ & $+$ & $-$ \\
          $10_3$  &  $t_{3}$  &  $j$    & $+$ & $+$ & $-$ & $-$ & $+$  
& $+$ & $-$ & $-$ \\
          $10_5$  &   $t_5$   &  $k$    & $+$ & $+$ & $+$ & $+$ & $-$  
& $-$ & $-$ & $-$ \\
         $5_{1}$  &  $-2t_1$  &  $jk$   & $+$ & $+$ & $-$ & $-$ & $-$  
& $-$ & $+$ & $+$ \\
         $5_{13}$ & $-t_1-t_3$& $ + $   & $+$ & $+$ & $+$ & $+$ & $+$  
& $+$ & $+$ & $+$ \\
         $5_{15}$ & $-t_1-t_5$&  $i$    & $+$ & $-$ & $+$ & $-$ & $+$  
& $-$ & $+$ & $-$ \\
         $5_{35}$ & $-t_3-t_5$&  $j$    & $+$ & $+$ & $-$ & $-$ & $+$  
& $+$ & $-$ & $-$ \\
         $5_{3}$  &  $-2t_3$  &  $-j$  & $-$ & $-$ & $+$ & $+$ & $-$ &  
$-$ & $+$ & $+$\\\hline
\end{tabular}
\end{center}
\caption{All possible matter parity assignments\label{tab:allpar}}
\end{table}

As such, models from $Z_2 \times Z_2$ are completely specified by the  
information present in table \ref{tab:modelbuilding}.

\begin{table}[h]
\begin{center}
\begin{tabular}{|c|c|c|c|}\hline
          Curve   &   Charge   & Matter Parity &                        
    Spectrum                           \\ \hline
          $10_1$  &  $t_{1}$   &     $i        $     &       $M_{10_1}  
Q+(M_{10_1}-N_1)u^c+(M_{10_1}+N_1)e^c    $        \\
          $10_3$  &  $t_{3}$   &      $j$      &       $M_{10_3}  
Q+(M_{10_3}-N_2)u^c+(M_{10_3}+N_2)e^c    $        \\
          $10_5$  &   $t_5$    &      $k$      &   $M_{10_5}  
Q+(M_{10_5}+N_1+N_2)u^c+(M_{10_5}-N_1-N_2)e^c    $    \\
         $5_{1}$  &  $-2t_1$   &     $jk$      &       $M_{5_1}  
\overline{d^c} + (M_{5_1}-N_1) \overline{L}$       \\
         $5_{13}$ & $-t_1-t_3$ &     $ + $     &    $M_{5_{13}}  
\overline{d^c} + (M_{5_{13}}+2 N_1) \overline{L}$    \\
         $5_{15}$ & $-t_1-t_5$ &      $i$      &    $M_{5_{15}}  
\overline{d^c} + (M_{5_{15}}+N_1) \overline{L}$    \\
         $5_{35}$ & $-t_3-t_5$ &     $j$      & $M_{5_{35}}  
\overline{d^c} + (M_{5_{35}}-2 N_1-N_2) \overline{L}$ \\
         $5_{3}$  &  $-2t_3$   &     $-j$      &     $M_{5_{3}}  
\overline{d^c} + (M_{5_{3}}+N_2) \overline{L}$\\\hline
\end{tabular}
\end{center}
\caption{All the relevant information for model building with $Z_2  
\times Z_2$ monodromy\label{tab:modelbuilding}}
\end{table}

\subsection{The Singlets}

For the singlets on the GUT surface we start by looking at the  
splitting equation for singlet states, $P_0$. 
For $SU(5)$  these are found  to be
\al{
P_0                                &=3125 b_5^4 b_0^4+256 b_4^5  
b_0^3-3750 b_2 b_3 b_5^3 b_0^3+2000 b_2 b_4^2 b_5^2 b_0^3+2250 b_3^2  
b_4 b_5^2 b_0^3-1600 b_3 b_4^3 b_5 b_0^3-128 b_2^2 b_4^4 b_0^2          
\nonumber\\
                 &+144 b_2 b_3^2 b_4^3 b_0^2-27 b_3^4 b_4^2 b_0^2+825  
b_2^2 b_3^2 b_5^2 b_0^2-900
                 b_2^3 b_4 b_5^2 b_0^2+108 b_3^5 b_5   b_0^2+560 b_2^2  
b_3 b_4^2 b_5 b_0^2-630 b_2 b_3^3 b_4 b_5 b_0^2\nonumber\\
                 &+16 b_2^4 b_4^3 b_0-4 b_2^3 b_3^2 b_4^2 b_0+108  
b_2^5 b_5^2 b_0+16 b_2^3 b_3^3 b_5 b_0-72 b_2^4 b_3 b_4 b_5 b_0
}
Applying the solution for the $Z_2 \times Z_2$ monodromy  
 from Eq.(\ref{eq:anhomol},\ref{eq:splitsol}) the 
above splits into 13 factors as follows
\al{
P_0 &=a_6^2 a_8^2 c \left(a_5^2-4 a_4 a_6\right) \left(a_8 (a_4  
a_8-a_5 a_7)+a_6 a_7^2\right)^2 \nonumber 
\\
&\left(c (a_5 a_8+a_6 a_7)^2-4 a_1 a_6  
a_8\right)(a_1 a_8+a_7 c (a_5 a_8+2 a_6 a_7))^2 \nonumber \\
& \left(a_1^2 a_6+a_1 c \left(-2 a_4 a_6 a_8+2 a_5^2 a_8+a_5 a_6  
a_7\right)+a_4 c^2 \left(a_6 a_8 (a_4 a_8+3 a_5 a_7)+2 a_5^2  
a_8^2+a_6^2 a_7^2\right)\right)^2\label{eq:P0solved}
}
 Their homologies   and geometric parities can be founded by applying the results from the 
previous section, and are presented in Table~\ref{tab:singfull}

\begin{table}[h]
\begin{center}
\begin{tabular}{|c|c|c|c|c|}\hline
 Equation     &Power  &     Charge      &            Homology     Class         & Matter Parity 
 \\ \hline
 $a_6$   &2     & $\pm(t_1-t_3)$  &  $\chi_2$         & $j        $         
  \\
  $a_8$  &2      & $\pm(t_1-t_5) $ &       $\eta-\chi_1-\chi_2 $      & $-k$         
   \\
 $c$   &1 &     $0$       &          $-\eta+2\chi_1$         & $ijk$         \\
 $a_5^2-\ldots$  &1  &       $0$       &        $-2   c_1+\chi_2$         & $+$           \\
  $a_8(a_4a_8-\ldots$&2 & $\pm(t_3-t_5)$  & $2 \eta-2 c_1-2  \chi_1-\chi_2$ & $j$           \\
 $c(a_5a_8+\ldots$&1  &       $0$       &          $\eta-2 c_1$     & $ijk$         \\
 $(a_1a_8+\ldots$ &2  & $\pm(t_1-t_5) $ &      $  \eta-2c_1-\chi_2 $      & $ -ik $       \\
  $(a_1^2 a_6+\ldots$&2 & $\pm(t_1-t_3)$  &    $-4 c_1+2   \chi_1+\chi_2$    & $j$
\\\hline
\end{tabular}
\end{center}
\caption{Defining euations, multiplicity, homologies,  matter parity, 
and perpendicular charges of   singlet factors\label{tab:singfull}}
\end{table}

\subsection{Application of Geometric Matter Parity }
We study now the implementation of the explicit $Z_2 \times Z_2$  
monodromy model presented in~\cite{Dudas:2010zb} alongside the  
matter parity proposed above. The model under consideration is defined by the  
flux data
\al{
&N_1 =  M_{5_{15}} = M_{5_{35}}=0 \\
&N_2 = M_{10_3}= M_{5_1}=1=-M_{10_5}=-M_{5_3} \\
&M_{10_1} = 3 = - M_{5_{13}}
}
which leads to the spectrum presented in  \tref{tab:dudaspaltimodels}  
alongside all possible geometric parities.

\begin{table}[h]
\begin{center}
\begin{tabular}{|c|c|c|c|c|c|c|c|c|c|c|}\hline
          Curve   &   Charge   &              Spectrum              &  
\multicolumn{7}{c}{All possible assignments}&  \\ \hline
          $10_1$  &  $t_{1}$   &        $3 Q+3u^c+3e^c    $         &  
$+$ & $-$ & $+$ & $-$ & $+$ & $-$ & $+$ & $-$  \\ 
          $10_3$  &  $t_{3}$   &            $Q+2e^c    $            &  
$+$ & $+$ & $-$ & $-$ & $+$ & $+$ & $-$ & $-$  \\ 
          $10_5$  &   $t_5$    &           $- Q-2e^c    $           &  
$+$ & $+$ & $+$ & $+$ & $-$ & $-$ & $-$ & $-$  \\
         $5_{1}$  &  $-2t_1$   &   $D_u +H_u$   & $+$ & $+$ & $-$ &  
$-$ & $-$ & $-$ & $+$ & $+$  \\
         $5_{13}$ & $-t_1-t_3$ & $-3 \overline{d^c} -3\overline{L}$ &  
$+$ & $+$ & $+$ & $+$ & $+$ & $+$ & $+$ & $+$  \\
         $5_{15}$ & $-t_1-t_5$ &                $0$                 &  
$+$ & $-$ & $+$ & $-$ & $+$ & $-$ & $+$ & $-$  \\ 
         $5_{35}$ & $-t_3-t_5$ &          $- \overline{H}_d$           
& $+$ & $+$ & $-$ & $-$ & $+$ & $+$ & $-$ & $-$  \\ 
         $5_{3}$  &  $-2t_3$   &         $- \overline{D}_d$         &  
$-$ & $-$ & $+$ & $+$ & $-$ & $-$ & $+$ & $+$\\\hline
\end{tabular}
\end{center}
\caption{Spectrum and allowed geometric parities for the  
$Z_2\times Z_2$ monodromy model\label{tab:dudaspaltimodels}}
\end{table}

\begin{table}[h]
\begin{center}
\begin{tabular}{|c|c|c|c|c|c|c|c|c|c|}\hline
         Name        &     Charge            & \multicolumn{7}{c}{All  
possible assignments}& \\ \hline
         $\theta_1$ & $\pm(t_1-t_3)$        & $+$ & $+$ & $-$ & $-$ &  
$+$ & $+$ & $-$ & $-$     \\ \hline
         $\theta_2$ & $\pm(t_1-t_5) $              & $-$ & $-$ & $-$ &  
$-$ & $+$ & $+$ & $+$ & $+$     \\ \hline
         $\theta_3$ &       $0$             & $+$ & $-$ & $-$ & $+$ &  
$-$ & $+$ & $+$ & $-$    \\ \hline
         $\theta_4$ &       $0$                    & $+$ & $+$ & $+$ &  
$+$ & $+$ & $+$ & $+$ & $+$      \\ \hline
         $\theta_5$ & $\pm(t_3-t_5)$                           & $+$ &  
$+$ & $-$ & $-$ & $+$ & $+$ & $-$ & $-$      \\ \hline
         $\theta_6$ &       $0$             & $+$ & $-$ & $-$ & $+$ &  
$-$ & $+$ & $+$ & $-$    \\ \hline
         $\theta_7$ & $\pm(t_1-t_5) $       & $-$ & $+$ & $-$ & $+$ &  
$+$ & $-$ & $+$ & $-$  \\ \hline
         $\theta_8$ & $\pm(t_1-t_3)$                     & $+$ & $+$ &  
$-$ & $-$ & $+$ & $+$ & $-$ & $-$\\\hline
\end{tabular}
\end{center}
\caption{Singlet curves and their perpendicular charges and geometric  
parity\label{tab:singlets}}
\end{table}

Inspecting \tref{tab:dudaspaltimodels} one can arrive at some  
conclusions. For example, looking at the spectrum from each curve it's  
immediate that all matter is contained in $10_1$ and $5_{13}$, while  
the Higgses come from $5_1$ and $5_{35}$, and the rest of the states  
are exotics that come in vector-like pairs. Immediately we see that  
there will be R-Parity violating terms since $5_{13}$ has positive  
parity. 

In order to fully describe the model one also has to take into account  
the singlets, whose perpendicular charges and all possible geometric  
parities can be seen in  \tref{tab:singlets}, where we included  
the same field with its charge conjugated partner in the same row - i.e. $\theta_i$ has the same parity as $\overline{\theta}_i$. \\

Of the possible combinations $\{i,j,k\}$ for the geometric parity assignments, 
the only choices that allow for a tree-level top quark mass are:
\begin{align}
\{i,j,k\}=\{+,+,+\}\\
\{i,j,k\}=\{-,+,+\}\nonumber\\
\{i,j,k\}=\{+,-,-\}\nonumber\\
\{i,j,k\}=\{-,-,-\}\nonumber
\end{align}
The option that most closely resembles the  R-parity imposed in the  model~\cite{Dudas:2010zb}  corresponds 
to the choice $i=-$, $j=k=+$. However, if R-parity has a geometric origin
the parity assignments of matter curves cannot be arbitrarily chosen. Using the Mathematica package presented in \cite{Fonseca:2011sy}, it is straight forward to produce the spectrum of operators up to an arbitrary mass dimension.
One can readily observe that its implementation 
allows  a number of operators that could cause Bilinear R-Parity 
 Violation (BRPV) at unacceptably high rates. For example, the lowest order operators are:
\begin{align}
H_uL\theta_1,\, 
H_uL\theta_8,\, 
H_uL\theta_1\theta_4,\, 
H_uL\theta_4\theta_8,\, 
H_uL\overline{\theta}_5\theta_7 
\end{align}
with higher order operators also present, amplifying the scale of the problem. 
In order to avoid problems, we must forbid vacuum expectations for a number of singlets,
 especially $\theta_1$ and $\theta_8$. This does not immediately appear to be a
  model killing issue, however we must look to the exotic masses. 
  Considering the Higgs triplets $D_{u/d}$, the only mass terms are:
\begin{align}
D_uD_d\theta_1\theta_1\theta_3,\, 
D_uD_d\theta_1\theta_1\theta_6,\, 
D_uD_d\theta_1\theta_2\overline{\theta}_5,\, 
D_uD_d\theta_1\theta_3\theta_8,\,\nonumber \\
D_uD_d\theta_1\theta_6\theta_8,\, 
D_uD_d\theta_2\overline{\theta}_5\theta_8,\, 
D_uD_d\theta_3\theta_8\theta_8,\, 
D_uD_d\theta_6\theta_8\theta_8 
\end{align}
As can be seen each of these terms contains $\theta_1$ or $\theta_8$. 
Since these are required to have no vacuum expectation value, 
it follows that the Higgs triplets cannot become massive. 
Since this is a highly disfavoured feature, we must rule out this model. 

It transpires that in a similar way, all the models with this flux assignment must 
be ruled out when we apply this geometric parity. This is due to the tension between
 BRPV terms and exotic masses, which seem to always be at odds in models with this novel parity.
  This motivates one to search for models without any exotics, as these models will not have any
   constraining features coming from exotic masses, and we shall analyse one such model in the subsequence.

\section{Deriving the MSSM with the seesaw mechanism }\label{s:mod}

The parameter space of models is very large, given the number of reasonable combinations of fluxes, multiplicities and choices of geometric parities. There are a number of ways to narrow the parameter space of any search, for example requing  that there be no exotics present in the spectrum, or contriving there to be only one tree-level Yukawa (to enable a heavy top quark), or perhaps allowing only models with standard matter parity be considered. This last option is quite difficult to search for, but can be constructed. 

Let us make a choice for the flux parameters that enables this standard matter parity:
\begin{align}
&\{N_1=1,N_2=0\}\nonumber \\
&M_{10_1}=-M_{5_{13}}=2\nonumber\\
&M_{10_5}=-M_{5_3}=1\\
&M_{10_3}=M_{5_1}=M_{5_{13}}=M_{5_{35}}=0\nonumber\\
&i=-j=k=-\nonumber
\end{align}
The matter spectrum of this model is summarised in \tref{tab:model2}. With this choice,
 \tref{tab:singlets} will select the column with only the singlets $\theta_7$ and $\overline{\theta}_7$ 
 having a negative matter parity. Provided this singlet does not acquire a vacuum expectation
  it will then be impossible for Bilinear R-parity violating terms due to the nature of the parity 
  assignments. This will also conveniently give us candidates for right-handed neutrinos,
  $\theta_7$ and $\overline{\theta}_7$.  

\begin{table}[t!]
\begin{center}
\begin{tabular}{|c|c|c|c|}\hline
          Curve   &   Charge   & Matter Parity &                        
    Spectrum                           \\ \hline
          $10_1$  &  $t_{1}$   &     $-        $     &       $Q_3+Q_2+u^c_3+3e^c$        \\
          $10_3$  &  $t_{3}$   &      $+$      &       $ -  $        \\
          $10_5$  &   $t_5$    &      $-$      &   $Q_1+u^c_2+u^c_1  $    \\
         $5_{1}$  &  $-2t_1$   &     $-$      &       $-\overline{L}_1$       \\
         $5_{13}$ & $-t_1-t_3$ &     $ + $     &    $ 2H_u $    \\
         $5_{15}$ & $-t_1-t_5$ &      $-$      &    $-\overline{d}_2^c-\overline{d}_1^c - \overline{L}_2$    \\
         $5_{35}$ & $-t_3-t_5$ &     $+$      & $-2\overline{H}_d$ \\
         $5_{3}$  &  $-2t_3$   &     $-$      &     $-\overline{d}^c_3-\overline{L}_3$\\
$1_{15}=\theta_7$& $t_1-t_5$ & $-$ & $N_R^a$\\
$1_{51}=\overline{\theta}_7$& $t_5-t_1$ & $-$ & $N_R^b$\\\hline
\end{tabular}
\end{center}
\caption{Matter content for a model with the standard matter parity arising from a geometric parity assignment.\label{tab:model2}}
\end{table}

\subsection{Yukawas}
Having written down a spectrum that has the phenomenologically preferred R-parity, we must now examine the allowed couplings of the model. The model only allows Yukawa couplings to arise at non-renormalisable levels, however the resulting couplings give rise to rank three mass matrices. This is because the perpendicular group charges must be canceled out in any Yukawa couplings. For example, the Yukawa arising from $10_1\cdot10_1\cdot5_{13}$ has a charge $t_1-t_3$, which may be canceled by the $\theta_{1/8}$ singlets. Consider the Yukawas of the Top sector,
\begin{align}
10_1\cdot10_1\cdot5_{13}\cdot(\overline{\theta}_1+\overline{\theta}_8)&\rightarrow (Q_3+Q_2) u_3 H_u(\overline{\theta}_1+\overline{\theta}_8) \nonumber\\
10_1\cdot10_5\cdot5_{13}\cdot\theta_5&\rightarrow ((Q_3+Q_2) (u_1+u_2)+ Q_1 u_3) H_u \theta_5\\ 
10_5\cdot10_5\cdot5_{13}\cdot\theta_2\cdot\theta_5&\rightarrow Q_1 (u_1+u_2) H_u\theta_2\theta_5 \nonumber
\end{align}
where the numbers indicate generations (1, 2 and 3). The resulting mass matrix should be rank three, however the terms will not all be created equally and the rank theorem \cite{Cecotti:2009zf} should lead to suppression of operators arising from the same matter curve combination:
\begin{equation} 
M_{u,c,t}\sim v_u \left(\begin{array}{ccc}
\epsilon\theta_2\theta_5&\theta_2\theta_5&\theta_5\\
\epsilon^2\theta_5&\epsilon\theta_5&\epsilon(\overline{\theta}_1+\overline{\theta}_8)\\
\epsilon\theta_5&\theta_5&\overline{\theta}_1+\overline{\theta}_8
\end{array}\right)
\end{equation}
where each element of the matrix has some arbitrary coupling constant. We use here $\epsilon$ to denote suppression due to the effects of the Rank Theorem \cite{Cecotti:2009zf} for Yukawas arising from the same GUT operators. The lightest generation will have the lightest mass due to an extra GUT scale suppression arising from the second singlet involved in the Yukawa. There are a large number of corrections at higher orders in singlet VEVs, which we have not included here for brevity. These corrections will also be less significant compared to the lowest order contributions. 

In a similar way, the Down-type Yukawa couplings arise as non-renormalisable operators, coming from four different combinations. The operators for this sector often exploit the tracelessness of \su{5}, so that the sum of the GUT charges must vanish. The leading order Yukawa operators,
\begin{align}
10_1\cdot\overline{5}_{3}\cdot\overline{5}_{35}\cdot(\theta_1+\theta_8)&\rightarrow (Q_3+Q_2)d_3 H_d(\theta_1+\theta_8) \nonumber\\ 
10_1\cdot\overline{5}_{15}\cdot\overline{5}_{35}\cdot\theta_5&\rightarrow (Q_3 +Q_2) (d_1+d_2) H_d \theta_5\\ 
10_5\cdot\overline{5}_{3}\cdot\overline{5}_{35}\cdot(\theta_1+\theta_8)\theta_2 &\rightarrow Q_1 d_3H_u(\theta_1+\theta_8)\theta_2 \nonumber\\
10_5\cdot\overline{5}_{15}\cdot\overline{5}_{35}\cdot \theta_2\cdot\theta_5&\rightarrow Q_1 (d_1+d_2) H_u\theta_2\theta_5 \nonumber
\end{align}
The resulting mass matrix will, like in the Top sector, be a rank three matrix, with a similar form:
\begin{equation} 
M_{d,s,b}\sim v_d \left(\begin{array}{ccc}
\epsilon\theta_2\theta_5&\theta_2\theta_5&(\theta_1+\theta_8)\theta_2\\
\epsilon^2\theta_5&\epsilon\theta_5&\epsilon(\theta_1+\theta_8)\\
\epsilon\theta_5&\theta_5&\theta_1+\theta_8
\end{array}\right)
\end{equation}
The structure of the Top and Bottom sectors appears to be quite similar in this model, which should provide a suitable hierarchy to both sectors.

The Charged Leptons will have a different structure to the Bottom-type quarks in this model, due primarily to the fact the $e^c_i$ matter is localised on one GUT tenplet. The Lepton doublets however all reside on different $\overline{5}$ representations, which will fill out the matrix in a non-trivial way, with the operators:
\begin{align}
10_1\cdot\overline{5}_{3}\cdot\overline{5}_{35}\cdot(\overline{\theta}_1+\overline{\theta}_8)&\rightarrow  L_3(e^c_1+e^c_2+e^c_3)H_d (\overline{\theta}_1+\overline{\theta}_8)\nonumber\\
10_1\cdot\overline{5}_{15}\cdot\overline{5}_{35}\cdot\theta_5&\rightarrow L_2(e^c_1+e^c_2+e^c_3)H_d \theta_5 \\
10_1\cdot\overline{5}_{1}\cdot\overline{5}_{35}\cdot (\theta_1+\theta_8)&\rightarrow L_1(e^c_1+e^c_2+e^c_3)H_d (\theta_1+\theta_8) \nonumber
\end{align}
The mass matrix for the Charged Lepton sector will be subject to suppressions arising due to the effects discussed above.

\subsection{Neutrino Masses}
The spectrum contains two singlets that do not have vacuum expectation values, which protects the model from certain classes of  dangerous operators. These singlets,  $\theta_7/\overline{\theta}_7$, also serve as candidates for right-handed neutrinos. Let us make the assignment $\theta_7=N_R^a$ and $\overline{\theta}_7=N_R^b$. This gives Dirac masses from two sources, the first of which involve all lepton doublets and $N_R^a$:
\begin{align}
\overline{5}_{3}\cdot5_{13}\cdot\theta_7\cdot \overline{\theta}_5&\rightarrow L_3N_R^aH_u\overline{\theta}_5   \nonumber\\
\overline{5}_{15}\cdot5_{13}\cdot\theta_7\cdot (\overline{\theta}_1+\overline{\theta}_8)&\rightarrow L_2N_R^aH_u (\overline{\theta}_1+\overline{\theta}_8) \\
\overline{5}_{1}\cdot5_{13}\cdot \theta_7\cdot (\overline{\theta}_1+\overline{\theta}_8)\cdot \theta_2&\rightarrow L_1N_R^aH_u (\overline{\theta}_1+\overline{\theta}_8)\theta_2\nonumber
\end{align}
This generates a hierarchy for neutrinos, however the effect will be mitigated by the operators arising from the $N_R^b$ singlet:
\begin{align}
\overline{5}_{3}\cdot5_{13}\cdot\overline{\theta}_7\cdot (\overline{\theta}_1+\overline{\theta}_8)\cdot \theta_2&\rightarrow L_3N_R^b H_u (\overline{\theta}_1+\overline{\theta}_8)\theta_2 \nonumber\\
\overline{5}_{15}\cdot5_{13}\cdot\overline{\theta}_7\cdot \theta_2\cdot \theta_5&\rightarrow L_2N_R^b H_u\theta_2 \theta_5\\
\overline{5}_{1}\cdot5_{13}\cdot \overline{\theta}_7\cdot \theta_5&\rightarrow L_1N_R^b H_u\theta_5 \nonumber
\end{align}
If all these Dirac mass operators are present in the low energy spectrum, then the neutrino sector should have masses that mix greatly. This is compatible with our understanding of neutrinos from experiments, which requires large mixing angles compared to the other sectors. 

A light mass scale for the neutrinos can be generated using the seesaw mechanism
\cite{seesaw}, which requires large right-handed Majorana masses to generate light physical left-handed Majorana
neutrino mass at low values. The singlets involved in this scenario has perpendicular charges that must be canceled out, as with the quark and charged lepton operators. Fortunately, this can be achieved, in part due to the presence of $\theta_2/\overline{\theta}_2$, which have the same charge combinations as $N_R^{a,b}$. The leading contribution to the mass term will come from the off diagonal $\theta_7\overline{\theta}_7$ term, however there are diagonal contributions:
\begin{align}
 \frac{\langle\theta_2\rangle^2}{\Lambda}\overline{\theta}_7^2\,\,+\,\,\frac{\langle\overline{\theta}_2\rangle^2}{\Lambda}\theta_7^2\,\,+\,\,M\theta_7\overline{\theta}_7
\end{align}
Two right-handed neutrinos are sufficient to generate the appropriate physical light masses for the neutrinos required by experimental constraints \cite{King:1999mb,King:2015dvf}.

\subsection{Other Features }
An interesting property of this model is the requirement of extra Higgs fields. Due to the flux factors, under doublet-triplet splitting it is necessary to have two copies of the up and down-type Higgs. This  insures that the model is free of Higgs colour triplets, $D_u/D_d$ in the massless spectrum, while also allowing the designation of $+$ parity to Higgs matter curves.  As a consequence of this, the $\mu$-term for the Higgs mass would seem to give four Higgs operators of the same mass: $M_{ij}H_u^iH_d^j$, with $i,j=1,2$. However, since for both the up and down-types there are two copies on the matter curve, we can call upon the rank theorem \cite{Cecotti:2009zf}. Consider the operator for the $\mu$-term:
\begin{equation}
5_{13}\cdot\overline{5}_{35}\cdot \theta_2\rightarrow M_{ij}H_u^iH_d^j \rightarrow
M \left(\begin{array}{cc}
\epsilon_h^2 & \epsilon_h \\
\epsilon_h  & 1
\end{array}\right)\left(\begin{array}{c}H_u^1 \\ H_u^2\end{array}\right)\left(\begin{array}{cc}H_d^1&H_d^2\end{array}\right)
\end{equation}
This operator will give a mass that is naturally large for one generation of the Higgs, while the second mass should be suppressed due to non-perturbative effects. This is parameterised by $\epsilon_h$, which is required to be sufficiently small as to allow a Higgs to be present at the electroweak scale, while the leading order Higgsl must be heavy enough to remain at a reasonably high scale and not prevent unification. Thus we should have a light Higgs boson as well as a heavier copy that is as of yet undetected. 

The spectrum is free of the Higgs colour triplets $D_u/D_d$, however we must still consider operators of the types $QQQL$ and $d^cu^cu^ce^c$, since the colour triplets may appear in the spectrum at the string scale. Of these types of operator, most are forbidden at leading order due to the charges of the perpendicular group. However, one operator is allowed and we must consider this process:
\begin{equation}
10_110_110_5\overline{5}_3\rightarrow (Q_3+Q_2)(Q_3+Q_2)Q_1L_3+(u^c_2+u^c_1)u_3^cd^c_3(e^c_1+e^c_2+e^c_3)
\end{equation}
None of the operators arising are solely first generation matter, however due to mixing they may contribute to any proton decay rate. The model in question only has one of each type of Higgs matter curve, which means any colour triplet partners must respect the perpendicular charges of those curves. The result of this requirement is that the vertex between the initial quarks and the $D_u$ colour triplet must also include a singlet to balance the charge, with the same requirement for the final vertex. The resulting operator should be suppressed by some high scale where the colour triplets are appearing in the spectrum - $\Lambda_s$. The most dangerous contribution of this operator can be assume to be the $Q_2Q_1Q_2L_3$ component, which will mix most strongly with the lightest generation. It can be estimated that, given the quark mixing and the mixing structure of the charged Leptons in particular, the suppression scale should be in the region $\sim10^{4-6}\Lambda_s$. This estimate seems to place the suppression of proton decay at too small a value, though not wildly inconsistent. 

\begin{figure}[t!]\centering
\includegraphics[scale=0.6]{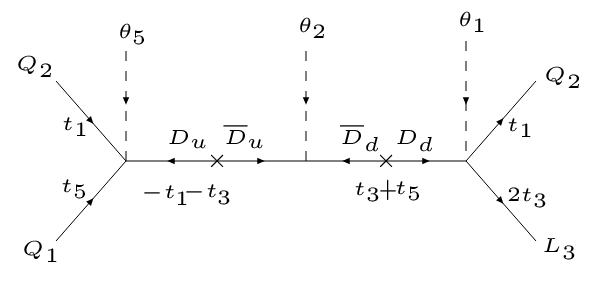}\label{pdpro}
\caption{Proton Decay graph}
\end{figure}

However, if we consider \fref{pdpro}, we can see that while the external legs of this process give an overall adherence to the charges of the perpendicular group charges, the vertices require singlet contributions. For example, the first vertex is $Q_2Q_1D_u\theta_5$, which is nonrenormalisable and  we cannot write down a series of renormalisable operators to mediate this effective operator. This is because the combination of perpendicular group and GUT charges constrain heavily the operators we can write down, which means proton decay can be seen to be suppressed here by the dynamics as well as the symmetries required by the F-theory formalism. The full determination of the coupling strengths of any process of this type in F-theory should be found through computing the overlap integral of the wavefunctions involved \cite{Camara:2011nj}, and this will be discussed in upcoming work on R-parity violating processes. 

\section{Conclusions}
We have revisited a class of $SU(5)$ SUSY GUT models which 
arise in the context of the spectral cover with 
Klein Group monodromy  $V_4=Z_2\times Z_2$.
By investigating the symmetry structures of the spectral cover equation and the
defining equations of the mater curves it is possible to understand the F-theory geometric origin of matter
parity, which has hitherto been just assumed in an {\it ad hoc } way.
In particular, we have shown how the simplest $Z_2$ matter parities
can be realised via the new geometric symmetries respected
by the spectral cover. 
By exploiting the various ways that these symmetries can be assigned,  
there are a large number of possible variants.

We have identified a rather minimal example of this kind,
where the low energy effective theory below the GUT scale is just the MSSM with no exotics and 
standard matter parity. Furthermore, by deriving general properties of the singlet sector, 
consistent with string vacua, including the D and 
F-flatness conditions,
we were able to identify two singlets,  which provide suitable candidates for a two right-handed neutrinos.
We were thus able to derive the MSSM extended by a two right-handed 
neutrino seesaw mechanism.
We also computed all baryon and lepton number violating operators
emerging from higher  non-renormalisable operators and found all dangerous operators to be forbidden.

\vspace{0.1in}
SFK acknowledges partial support from the STFC Consolidated ST/J000396/1 grant and 
the European Union FP7 ITN-INVISIBLES (Marie Curie Actions, PITN-
GA-2011-289442). AKM is supported by an STFC studentship. MCR acknowledges support from the FCT under
the grant SFRH/BD/84234/2012.
\vspace{0.1in}

\newpage
\appendix
\section{F-Flatness}
The super potential for the singlets involves a total of thirteen fields, which couple in such as way as to cancel all their perpendicular group charges and to have consistent parity.
\begin{align}
W\supset& c_{17} \theta_1 \overline\theta_1 \theta_3+c_{18} \theta_1 \overline\theta_1 \theta_4+c_{19} \theta_1
   \overline\theta_1 \theta_6+c_4 \theta_1 \overline\theta_1+c_{20} \theta_1 \overline\theta_2
   \theta_5+ c_{21} \theta_1 \theta_3 \overline\theta_8+\nonumber\\
     & c_{22} \theta_1 \theta_4
   \overline\theta_8+c_{23} \theta_1 \theta_6 \overline\theta_8+c_5 \theta_1 \overline\theta_8+c_{24}
   \overline\theta_1 \theta_2 \overline\theta_5+c_{25} \overline\theta_1 \theta_3 \theta_8+c_{26} \overline\theta_1
   \theta_4 \theta_8+\nonumber\\
     & c_{27} \overline\theta_1 \theta_6 \theta_8+c_6 \overline\theta_1 \theta_8+ c_{28}
   \theta_2 \overline\theta_2 \theta_3+c_{29} \theta_2 \overline\theta_2 \theta_4+c_{30} \theta_2
   \overline\theta_2 \theta_6+c_7 \theta_2 \overline\theta_2+\nonumber\\
     & c_{31} \theta_2 \overline\theta_5
   \overline\theta_8+c_{32} \overline\theta_2 \theta_5 \theta_8+c_{33} \theta_3^3+c_{34} \theta_3^2
   \theta_4+c_{35} \theta_3^2 \theta_6+c_8 \theta_3^2+c_{36} \theta_3
   \theta_4^2+\nonumber\\
     & c_{37} \theta_3 \theta_4 \theta_6+c_9 \theta_3 \theta_4+c_{38} \theta_3
   \theta_5 \overline\theta_5+c_{39} \theta_3 \theta_6^2+c_{10} \theta_3 \theta_6+c_{40}
   \theta_3 \theta_7 \overline\theta_7+\nonumber\\
     & c_{41} \theta_3 \theta_8 \overline\theta_8+c_1 \theta_3+c_{42}
   \theta_4^3+c_{43} \theta_4^2 \theta_6+c_{11} \theta_4^2+c_{44} \theta_4 \theta_5
   \overline\theta_5+c_{45} \theta_4 \theta_6^2+\nonumber\\
     & c_{12} \theta_4 \theta_6+c_{46} \theta_4
   \theta_7 \overline\theta_7+c_{47} \theta_4 \theta_8 \overline\theta_8+c_2 \theta_4+c_{48} \theta_5
   \overline\theta_5 \theta_6+c_{13} \theta_5 \overline\theta_5+c_{49} \theta_6^3+\nonumber\\
     & c_{14}
   \theta_6^2+c_{50} \theta_6 \theta_7 \overline\theta_7+c_{51} \theta_6 \theta_8
   \overline\theta_8+c_3 \theta_6+c_{15} \theta_7 \overline\theta_7+c_{16} \theta_8 \overline\theta_8
\end{align}
In order to establish flatness of the F-terms, we must consider $F_{\theta_i}=\frac{\delta W}{\delta\theta_i}$, giving a total of thirteen equations, each of which must vanish. The solution of this system of equations should be such that none of the coefficients are required to take special values to be natural and free of fine tuning.

\begin{align}
	F_{\theta_1}= & c_{17} \overline\theta_1 \theta_3+c_{18} \overline\theta_1 \theta_4+c_{19} \overline\theta_1 \theta_6+c_4 \overline\theta_1+c_{20} \overline\theta_2 \theta_5+c_{21} \theta_3 \overline\theta_8+c_{22} \theta_4 \overline\theta_8\nonumber\\
	     & +c_{23} \theta_6 \overline\theta_8+c_5 \overline\theta_8
    \\
	F_{\overline\theta_1}= & c_{17} \theta_1 \theta_3+c_{18} \theta_1 \theta_4+c_{19} \theta_1 \theta_6+c_4 \theta_1+c_{24} \theta_2 \overline\theta_5+c_{25} \theta_3 \theta_8\nonumber\\
	     & +c_{26} \theta_4 \theta_8+c_{27} \theta_6 \theta_8+c_6 \theta_8
    \\
	F_{\theta_2}= & c_{24} \overline\theta_1 \overline\theta_5+c_{28} \overline\theta_2 \theta_3+c_{29} \overline\theta_2 \theta_4+c_{30}\overline\theta_2 \theta_6+c_7 \overline\theta_2+c_{31} \overline\theta_5 \overline\theta_8                                                                          
    \\
	F_{\overline\theta_2} = &c_{20} \theta_1 \theta_5+c_{28} \theta_2 \theta_3+c_{29} \theta_2 \theta_4+c_{30}\theta_2 \theta_6+c_7 \theta_2+c_{32} \theta_5 \theta_8                                                                                                                                                                                                                                                                      
	\\
	F_{\theta_3} = & c_{17} \theta_1 \overline\theta_1+c_{21} \theta_1 \overline\theta_8+c_{25} \overline\theta_1 \theta_8+c_{28} \theta_2 \overline\theta_2+3 c_{33} \theta_3^2+2 c_{34} \theta_3 \theta_4 \nonumber\\
	     & +2 c_{35} \theta_3 \theta_6+2 c_8 \theta_3+c_{36} \theta_4^2+c_{37} \theta_4 \theta_6+c_9 \theta_4+c_{38} \theta_5 \overline\theta_5\nonumber\\
	          & +c_{39} \theta_6^2+c_{10} \theta_6+c_{40} \theta_7 \overline\theta_7+c_{41} \theta_8 \overline\theta_8+c_1
    \\
	 F_{\theta_4} = & c_{18} \theta_1 \overline\theta_1+c_{22} \theta_1 \overline\theta_8+c_{26} \overline\theta_1 \theta_8+c_{29} \theta_2 \overline\theta_2+c_{34} \theta_3^2+2 c_{36} \theta_3 \theta_4\nonumber\\
	      & +c_{37} \theta_3 \theta_6+c_9 \theta_3+3 c_{42} \theta_4^2+2 c_{43} \theta_4 \theta_6+2 c_{11} \theta_4+c_{44} \theta_5 \overline\theta_5+c_{45} \theta_6^2\nonumber\\
	           & +c_{12} \theta_6+c_{46} \theta_7 \overline\theta_7+c_{47} \theta_8 \overline\theta_8+c_2
\end{align}

\begin{align}
	F_{\theta_5} = & c_{20} \theta_1 \overline\theta_2+c_{32} \overline\theta_2 \theta_8+c_{38} \theta_3 \overline\theta_5+c_{44} \theta_4 \overline\theta_5+c_{48} \overline\theta_5 \theta_6+c_{13} \overline\theta_5                                                                                                                                                                                                                                                              
	\\
	F_{\overline\theta_5} = & c_{24} \overline\theta_1 \theta_2+c_{31} \theta_2 \overline\theta_8+c_{38} \theta_3 \theta_5+c_{44} \theta_4 \theta_5+c_{48} \theta_5 \theta_6+c_{13} \theta_5
	\\
	F_{\theta_6} = & c_{19} \theta_1 \overline\theta_1+c_{23} \theta_1 \overline\theta_8+c_{27} \overline\theta_1 \theta_8+c_{30} \theta_2 \overline\theta_2+c_{35} \theta_3^2+c_{37} \theta_3 \theta_4\nonumber\\
	     & +2 c_{39} \theta_3 \theta_6+c_{10} \theta_3+c_{43} \theta_4^2+2 c_{45} \theta_4 \theta_6+c_{12} \theta_4+c_{48} \theta_5 \overline\theta_5+3 c_{49} \theta_6^2\nonumber\\
	          & +2 c_{14} \theta_6+c_{50} \theta_7 \overline\theta_7+c_{51} \theta_8 \overline\theta_8+c_3
	\\
	F_{\theta_7} = & c_{40} \theta_3 \overline\theta_7+c_{46} \theta_4 \overline\theta_7+c_{50} \theta_6 \overline\theta_7+c_{15} \overline\theta_7                                                                                  
    \\
	 F_{\overline\theta_7} = & c_{40} \theta_3 \theta_7+c_{46} \theta_4 \theta_7+c_{50} \theta_6 \theta_7+c_{15} \theta_7
    \\
	 F_{\theta_8} = & c_{25} \overline\theta_1 \theta_3+c_{26} \overline\theta_1 \theta_4+c_{27} \overline\theta_1 \theta_6+c_6 \overline\theta_1+c_{32} \overline\theta_2 \theta_5+c_{41} \theta_3 \overline\theta_8+c_{47} \theta_4 \overline\theta_8\nonumber\\
	      & +c_{51} \theta_6 \overline\theta_8+c_{16} \overline\theta_8                                                                                                                                                                                   
	 \\
 	F_{\overline\theta_8} =&	c_{21} \theta_1 \theta_3+c_{22} \theta_1 \theta_4+c_{23} \theta_1 \theta_6+c_5 \theta_1+c_{31} \theta_2 \overline\theta_5+c_{41} \theta_3 \theta_8\nonumber\\
 	     & +c_{47} \theta_4 \theta_8+c_{51} \theta_6 \theta_8+c_{16} \theta_8                                                                                                                                                                                         
\end{align}

The only constraint coming from the model discussed in Section \ref{s:mod} is that $\theta_7/\overline{\theta}_7$ not have a vacuum expectation to protect from dangerous operators. This requirement does not over constrain the equations or create any unsightly relations, however it also leaves a solution of the F-term alignment that is hard to write down in a concise manner due to the complexity of the equations involved.

\end{document}